\documentclass[]{aastex631}

\shorttitle{Temporal and spatial characteristics of hard X-ray sources}
\shortauthors{Shabalin et al.}

\usepackage{amsmath}

\begin{document}

\title{Temporal and spatial characteristics of hard X-ray sources in flare model with a vertical current sheet} 

\author[0000-0003-3938-0146]{Alexander N. Shabalin}
\affiliation{Ioffe Institute, Russian Academy of Sciences, Saint-Petersburg 194021, Russia}

\author[0000-0001-7892-093X]{Evgeniia P. Ovchinnikova}
\affiliation{Ioffe Institute, Russian Academy of Sciences, Saint-Petersburg 194021, Russia}

\author[0000-0002-6693-5613]{Yuri E. Charikov}
\affiliation{Ioffe Institute, Russian Academy of Sciences, Saint-Petersburg 194021, Russia}

\begin{abstract}

We analyzed changes in the height of the coronal hard X-ray (HXR) source for flares SOL2013-05-13T01:50 and SOL2013-05-13T15:51. Analysis of the Reuven Ramaty High Energy Solar Spectroscopic Imager data revealed the downward motion of the HXR source and the separation of the sources by energy and height. In the early stages of the flares, a negative correlation was found between the HXR source area in the corona and HXR flux. For the SOL2013-05-13T15:51 event, an increasing trend in the time delay spectra at the footpoints was obtained. For both events, the spectra of the time delays in the coronal HXR source showed a decreasing trend across the energies in certain flare phases. To interpret the observed phenomena, we considered a flare model of collapsing traps and calculated the distribution functions of accelerated electrons along the magnetic loop using a nonstationary relativistic kinetic equation. This approach considers betatron and Fermi first-order acceleration mechanisms. The increasing trend of the time delay spectra at the footpoints was explained by the high mirror ratio in the magnetic loop and betatron acceleration mechanism. The observed features in the spatial and temporal behavior of the HXR sources, such as the negative correlation between the HXR source area and HXR flux, can be interpreted by the collapsing trap model.

\end{abstract}


\section{Introduction} \label{sec:intro}

According to observations in the extreme ultraviolet (EUV) range and model calculations, magnetic field configurations with a vertical current sheet in the corona are possible in active regions of the solar atmosphere \citep{Warren2018, Gary2018, Jiang2018}. In the standard flare model CSHKP (Carmichael-Sturrock-Hirayama-Kopp-Pneuman), the primary acceleration of charged particles occurs in the regions of the cusp and current sheet \citep{Lin2000, ZharkovaAcc2011, Chen2020b} above the soft X-ray loops. The ejection of collisionless plasma from the magnetic reconnection region is accompanied by the transformation of magnetic loops initially extended in the direction of the current sheet to a configuration close to a dipole (magnetic relaxation). The magnetic loops that relax in this manner undergo longitudinal and transverse contraction, which leads to the further acceleration of electrons through betatron acceleration and Fermi first-order acceleration \citep{Somov1997}. Magnetic field dynamic results in spatial and temporal features of hard X-ray (HXR) and radio sources in flares. For example, the relaxing magnetic field in the cusp explains the position of the HXR source above the loop \citep{Masuda1994, Masuda1995} and its brightness \citep{Fletcher1998, Shabalin2022}. 

Analysis of the relative motion of X-ray sources at the footpoints of the flaring loop and looptop allows us to determine the regions of magnetic field reconnection and acceleration of charged particles \citep{Bogachev2005, Krucker2005, Somov2005}. In the works \citep{Sui2003, Sui2004, Liu2004,Liu2008, Veronig2006, Joshi2007}, an observed decrease in the height of the X-ray source in some solar flares was noted at an early stage of the HXR flux rise. \citet{Sui2004} noted a decrease in the height of the coronal HXR sources at the beginning of the flare for three M-class flares that occurred on April 14–16, 2002. To explain the downward motion of the HXR sources, the authors proposed two mechanisms: one is associated with the relaxation of magnetic fields in the cusp region formed during magnetic reconnection, and the other is related to a change in the type of magnetic field reconnection from slow to Petschek regime. The structure of the magnetic field in the cusp region and the evolution of this structure in the collapsing trap model (hereinafter, the CollTr model) allows us to expect an altitude change of the HXR source, which can be caused by loop shrinkage \citep{Somov1997}. In models, loop contraction is associated with the "unshearing" of magnetic loops during energy dissipation. This process results in the converging motion of the footpoints and downward movement of the X-ray looptop source \citep{Ji2007,Zhou2013}. In \citep{Li2005, Zaitsev2020}, the shrinking of the loop at 34 GHz is explained by a current decrease due to an increase in resistance caused by the plasma instability of Rayleigh-Taylor within the chromospheric footpoints of the magnetic loop.

In addition to changes in the height of the HXR sources, the relaxation process of the magnetic fields in the CollTr model can manifest itself in an affect the hardness of the X-ray and radio spectra, as well as the time delay spectra. Changes in the hardness of the energy spectrum in the X-ray range during solar flares have been extensively studied in many papers. These changes are the most commonly reported soft-hard-soft (SHS) patterns \citep{Parks1969, Kane1970, Benz1977, Brown1985, Lin1987, Fletcher2002}. In addition to the SHS pattern, a soft-hard-harder (SHH) \citep{Frost1971, Cliver1986, Kiplinger1995, Lysenko2019} and hard-soft-hard (HSH) \citep{Shao2009} patterns have also been observed. According to the authors, possible reasons for the formation of these patterns may include a change in the efficiency of the accelerator, the presence of several stages of acceleration, the effects of particle transfer in the loop (owing to the return current), and the influence of turbulent modes in the regions of acceleration and radiation. \citet{Grigis2004} showed that elementary flare bursts also exhibit SHS patterns. According to \citep{Battaglia2006}, an analysis of spatially separated sources at the top and footpoints showed that the SHS pattern was observed separately in the spectra of each source. The authors concluded that this SHS pattern was generated in the accelerator. The CollTr model assumes a change in the magnetic field configuration associated with the relaxation of the magnetic loops (collapse). This process can affect the formation of the spectrum of accelerated electrons and, hence, the dynamics of the X-ray spectrum.

The relaxation process of magnetic fields can manifest itself in the influence on the dynamic spectrum of the time delays of HXR radiation. \citet{Charikov2015} demonstrated that the time delay spectra (TD spectra) depend on the energy and pitch-angle distributions of the accelerated electrons and the parameters of the magnetic field and plasma. Typically, the analysis of TD spectra in the HXR range is performed using full-Sun observational data \citep{Bai1979, Aschwanden1995, Bespalov1987, Tsap2015a}. In the full-Sun dataset, researchers encounter superimposed HXR flux originating from the chromospheric and coronal regions of the flare arcade. However, the analysis of HXR fluxes from the local regions of magnetic loops on the millisecond timescale, which is preferable for calculating time delays, is complicated because of poor counting statistics.

This study aims to validate the CollTr model based on HXR radiation data and the features of its evolution in the region below the vertical current sheet. Such manifestations may include a decrease or absence of an increase in the height of the HXR sources, a negative correlation between the area of the coronal source and the HXR flux, and features in the energy spectra and TD spectra, particularly at the initial stage of flare development. The observed features of HXR radiation were substantiated by numerically solving the relativistic nonstationary kinetic equation for accelerated electrons propagating in flaring magnetic loops. In particular, we emphasize that in the simulations, in addition to the transport processes, Coulomb losses and diffusion, return current, and magnetic mirroring, the equation also considers the influence of the magnetic field dynamics (collapse) in the initial phase of the flare loop relaxation, which leads to betatron and Fermi first-order acceleration. Sections \ref{sec:obs}--\ref{sec:delays} present the results of the spatial and temporal analyses of the HXR fluxes for the solar events SOL2013-05-13T01:50 and SOL2013-05-13T15:51. The results of modeling the propagation of accelerated electrons in a magnetic loop and the radiation generated by them in the HXR range are presented in sections \ref{sec:modeling} and \ref{sec:modeling_res}. The results of the observations and simulations are discussed in Section \ref{sec:conclusion}.

\section{Spatial characteristics of flare hard X-rays} \label{sec:obs}
\begin{figure*}[ht!]
\centering
\includegraphics[trim=1mm 1mm 1mm 1mm, clip, width=0.8\textwidth]{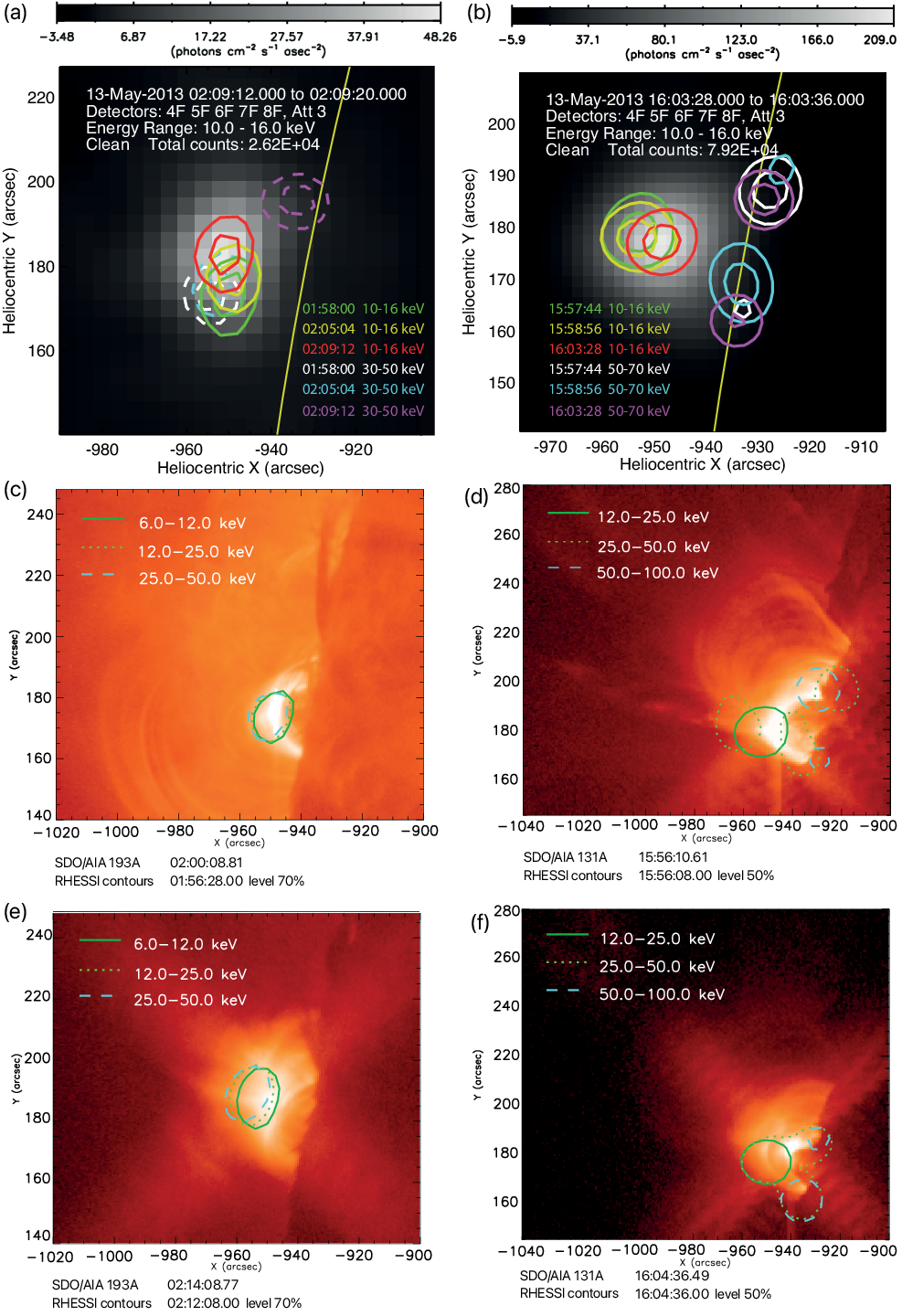}
\caption{RHESSI images at different times of the flares SOL2013-05-13T01:50 in panel (a) and SOL2013-05-13T15:51 in panel (b). SDO/AIA images of SOL2013-05-13T01:50 flare at 193 Å in the early phase and during decay (c,e). The same applies to panels (d) and (f) for SOL2013-05-13T15:51 flare at 131 Å. All the EUV images were superimposed by the RHESSI contours at different energy bands.}
\label{fig:images}
\end{figure*}

\begin{splitdeluxetable*}{lcccccBcccccc}
\caption{Correlation coefficients between the HXR light curves and areas of the HXR sources. Velocities of coronal HXR sources.
\label{tab:table1}}
\tablehead{
\colhead{} & \colhead{} & \multicolumn{4}{c}{Coefficient of correlation} &
\colhead{} & \colhead{} & \multicolumn{4}{c}{Velocity of coronal source, km s$^{-1}$}\\
\cline{3-6}
\cline{9-12}
\colhead{} & \colhead{} & \colhead{6-10 keV} & \colhead{10-16 keV} & \colhead{16-20 keV} & \colhead{20-25 keV} &\colhead{} &\colhead{} & \colhead{6-10 keV} & \colhead{10-16 keV} & \colhead{16-20 keV} & \colhead{20-25 keV}
}
\startdata 
{SOL2013-05-13T01:50} & {interval 1} & -0.5$\pm 0.2$ & -0.7$\pm 0.1$ & -0.7$\pm 0.1$ &  -- &
{SOL2013-05-13T01:50} & {interval 1} & -4.9          & -7.7          & -12.1         & -14.9 \\ 
{} & {interval 2} & -0.5$\pm 0.3$ & --   & -0.8$\pm 0.1$ &  -0.8$\pm 0.1$ &
{} & {interval 2} & 9.9           & 11.9 & 14.0          & 16.2 \\ 
\hline
{SOL2013-05-13T15:51}& {interval 1} & -0.5$\pm 0.3$ & -0.7$\pm 0.2$ & -0.6$\pm 0.3$ &  -- &
{SOL2013-05-13T15:51}& {interval 1} & 0.6           & -7.9          & -19.9         & -23.5\\ 
{ }& {interval 2} & +0.7$\pm 0.2$& +0.6$\pm 0.3$& +0.7$\pm 0.2$& +0.7$\pm 0.2$ &
{ }& {interval 2} & 0.6 & 9.1 & 3.1 & -1.3\\ 
{ }& {interval 3} & -0.4$\pm 0.3$& -0.8$\pm 0.1$& -0.6$\pm 0.2$&  -0.6$\pm 0.2$ &
{ }& {interval 3} & 11.6 & 14.8 & 17.4 & 22.1\\
\enddata
\tablecomments{The negative velocity values correspond to the downward motion of the coronal source.}
\end{splitdeluxetable*}

\begin{figure*}[ht!]
\gridline{\fig{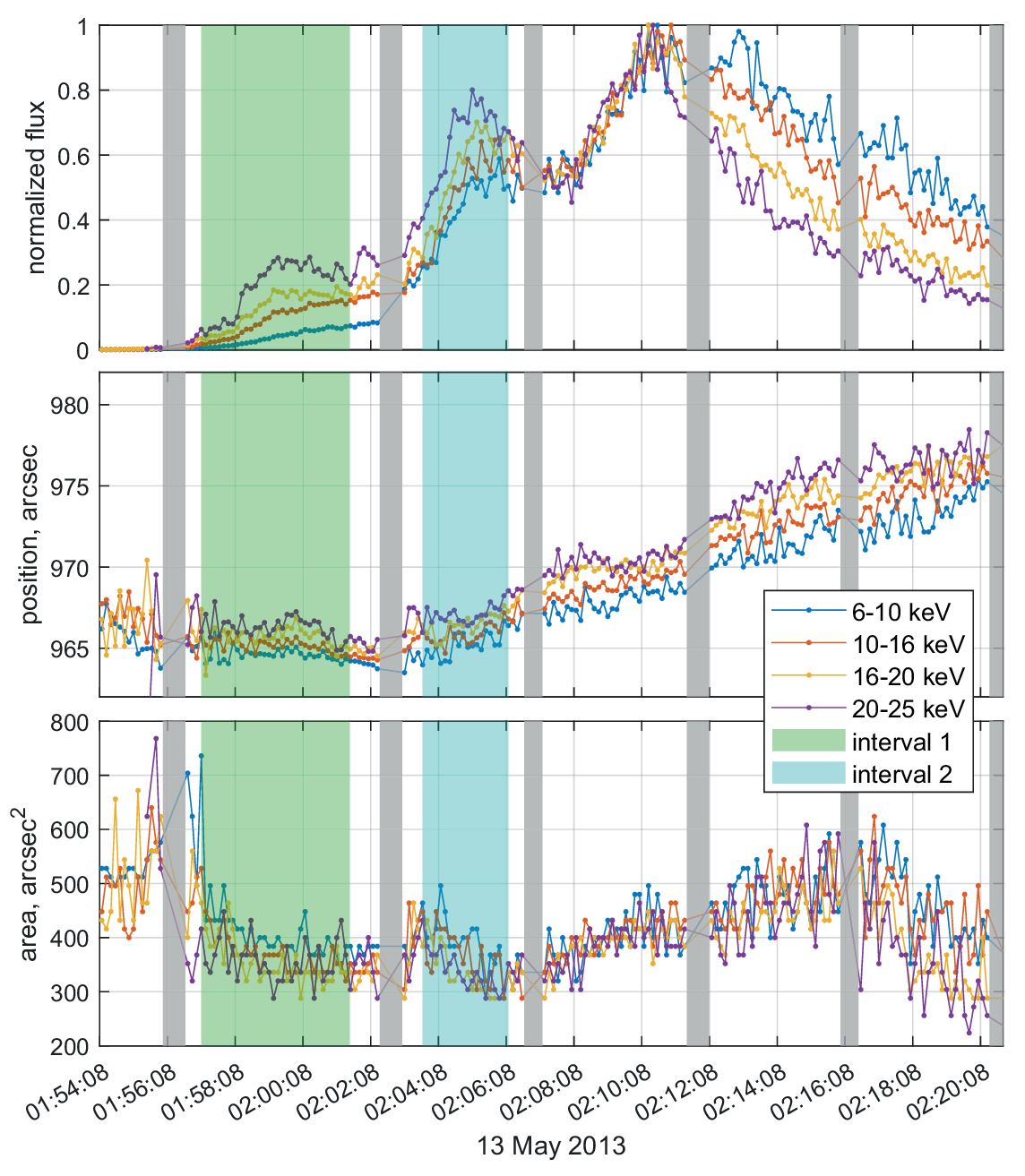}{0.45\textwidth}{(a)}
          \fig{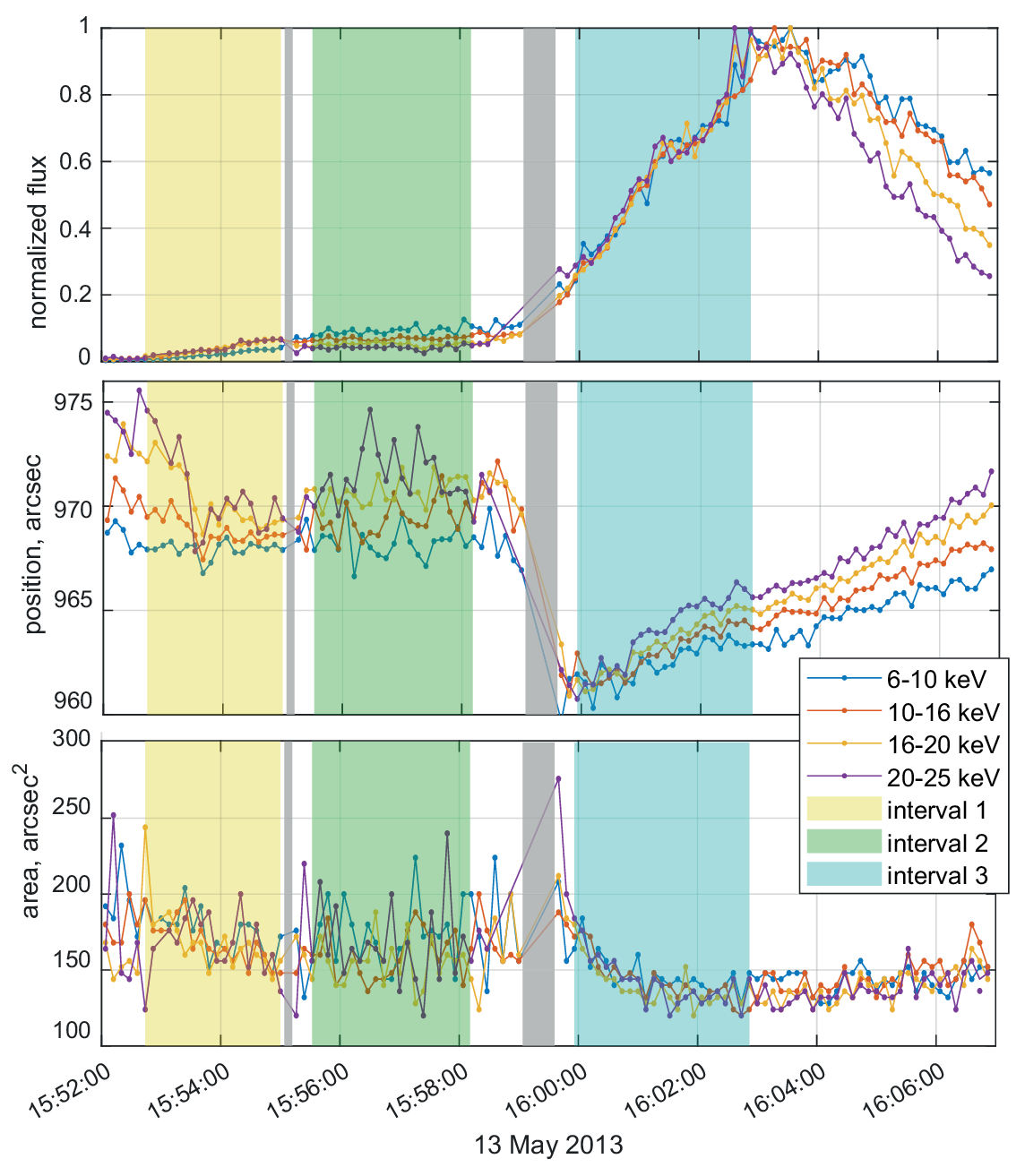}{0.45\textwidth}{(b)}
          }
\caption{RHESSI normalized time profiles, HXR source height, and area of HXR sources for (a) SOL2013-05-13T01:50, (b) SOL2013-05-13T15:51. The gray vertical stripes mark the intervals with a change in the attenuator state, excluded from the study.
\label{fig:Profiles}}
\end{figure*}

We present observations of the solar observatory, The Reuven Ramaty High Energy Solar Spectroscopic Imager (RHESSI) \citep{Lin2002}, of two X-class flares from the same active region, AR11748. The X-1.7 GOES class flare SOL2013-05-13T01:50 is partially occulted by a solar limb. During this flare, a single HXR source was observed in the corona prior to the main peak (Figure \ref{fig:images}a). Six minutes before flare onset, a precursor was observed in the extreme ultraviolet range and soft X-rays, which were studied in detail by \citep{Shen2017}. A second flare, SOL2013-05-13T15:51, X-2.8 GOES class, occurred 14 hours later. Three HXR sources were observed during this flare: two in the chromosphere and one in the corona (Figure \ref{fig:images}b). In both events, flare arcades were visible in ultraviolet data (Figure \ref{fig:images}c--\ref{fig:images}f) obtained with The Atmospheric Imaging Assembly on the Solar Dynamics Observatory SDO/AIA \citep{Lemen2012}. In panels \ref{fig:images}d and \ref{fig:images}e, the cusp and the thin structure above it can be observed in the EUV range. The long thin feature is considered as a possible current sheet position. The RHESSI contours in the energy bands of 6--12 and 12--25 keV coincide with the EUV cusp area in both flares. For flare SOL2013-05-13T15:51, the 50--100 keV contours coincided with the footpoints of the EUV loops (panels \ref{fig:images}b, \ref{fig:images}d, and \ref{fig:images}f). The images in panels \ref{fig:images}c and \ref{fig:images}d were obtained during the initial phase of the flux growth, whereas those in panels \ref{fig:images}e and \ref{fig:images}f were obtained during the decay phase.

RHESSI CLEAN images with an eight-second integration time were used to analyze the height and area of HXR sources. The 4F--8F detectors were utilized for image generation. The HXR source area was estimated as 50\% intensity contour for SOL2013-05-13T01:50 and 70\% intensity contour for SOL2013-05-13T15:51. The position of the HXR source was determined as the distance from the center of the solar disk to the HXR centroid in the plane of the sky. The time evolutions of the flux, height, and area of the coronal HXR source are shown in Figure \ref{fig:Profiles}.

During the initial phase of the SOL2013-05-13T01:50 flare, the altitude of the HXR source decreased by 1 arcsec over a period of 4 minutes. At the subsequent time interval, the height of the HXR source increased by 2 arcsec over a period of 3 minutes, and the HXR flux significantly increased. The average HXR source velocities for different energy bands are listed in Table \ref{tab:table1}. The velocities of the HXR sources in the downward and upward directions increased with energy. Negative velocity values corresponded to a decrease in the altitude of the coronal HXR sources, whereas positive values corresponded to a height increase. It is worth noting that the centroids of the HXR sources with higher energies are located higher than those of the HXR sources with lower energies (Figure \ref{fig:Profiles}a). 

The obtained HXR area values were subjected to correlation analyses with the HXR flux in the energy ranges 6–10, 10–16, 16–20, and 20–25 keV. Cross-correlation analysis of the coronal HXR source area and the HXR flux revealed a negative correlation. This was observed in interval 1 with decreasing height and in the subsequent interval 2 with increasing height of the HXR sources. The correlation coefficients are listed in Table \ref{tab:table1}.

In the SOL2013-05-13T15:51 flare, the change in the height of the HXR source exhibited step-like behavior (Figure \ref{fig:Profiles}b, middle panel). At the onset of the flare, there was a slight increase in the HXR flux, which coincided with a decrease in the height of the HXR source. In addition, the area of the HXR source showed an inverse correlation with the HXR flux increment. After the flare onset, the increase in the HXR flux ceases and is replaced by relatively stable radiation over a period of approximately 4 minutes. During the initial time phases at intervals 1 and 2 (Figure 2, middle panel on the right), the area values of the HXR sources displayed substantial fluctuations around a mean value of approximately 170 arcsec$^2$. The area of the HXR source during second time interval was positively correlated with the HXR flux, whereas the height of the HXR source above the solar limb did not exhibit a clear tendency to increase or decrease. This was followed by a rapid increase in the HXR flux (interval 3). At the beginning of this phase, within the interval 15:59--16:00 UT, the attenuators passed a switch. After the attenuators were switched, a sharp change in the source height was observed. It is possible that a brighter HXR source emerged, which was located approximately 8 arcsec lower. The emergence of this HXR source likely occurred during the transition of the flare to its explosive phase, and the location of this source differed within the considered magnetic configuration. A subsequent increase in the HXR flux was accompanied by an increase in the height of the HXR source above the solar limb. In this phase, the area of the HXR source showed an inverse correlation with the increase in the HXR flux. Observation of the flare revealed the separation of the HXR source centroids at various energy ranges throughout the event. The HXR sources observed at higher energies corresponded to greater heights above the solar limb, as shown in the middle panels of Figure \ref{fig:Profiles}b. It should be noted that the velocities of the downward and upward motions differed for different energy channels and had higher values at higher energies, as in the SOL2013-05-13T01:50 flare.

Thus, a negative correlation between the area of the corresponding HXR source and the increase in HXR flux was observed during the time intervals with an increase in the HXR flux in the considered flares. A positive correlation between the area and HXR flux corresponded to the time interval in which the area and HXR flux did not change significantly (Table \ref{tab:table1}).

\begin{figure*}[ht!]
\gridline{\fig{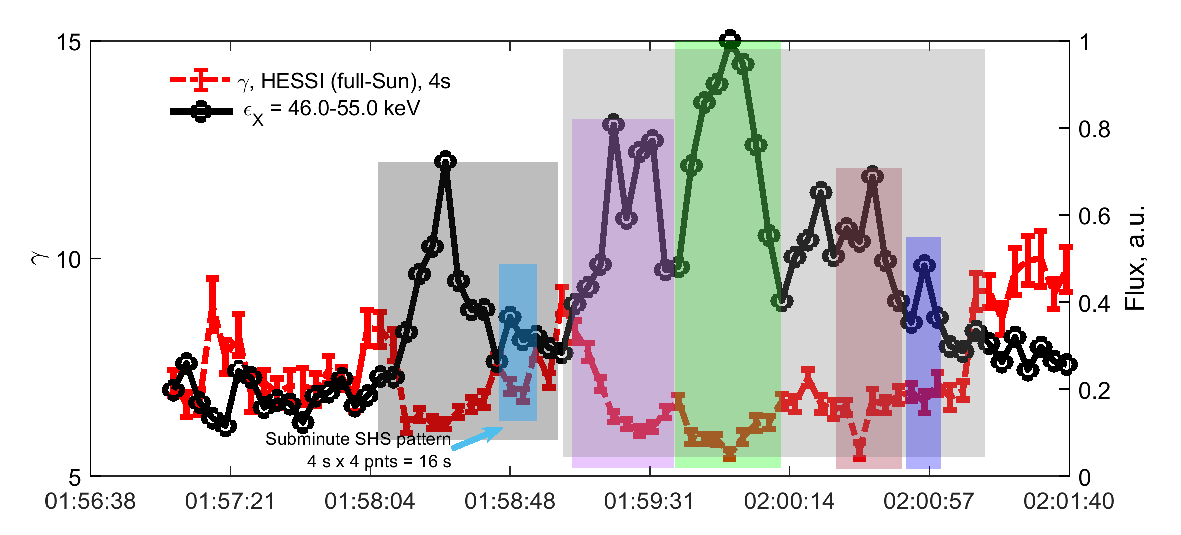}{0.95\textwidth}{}
    }
\gridline{\fig{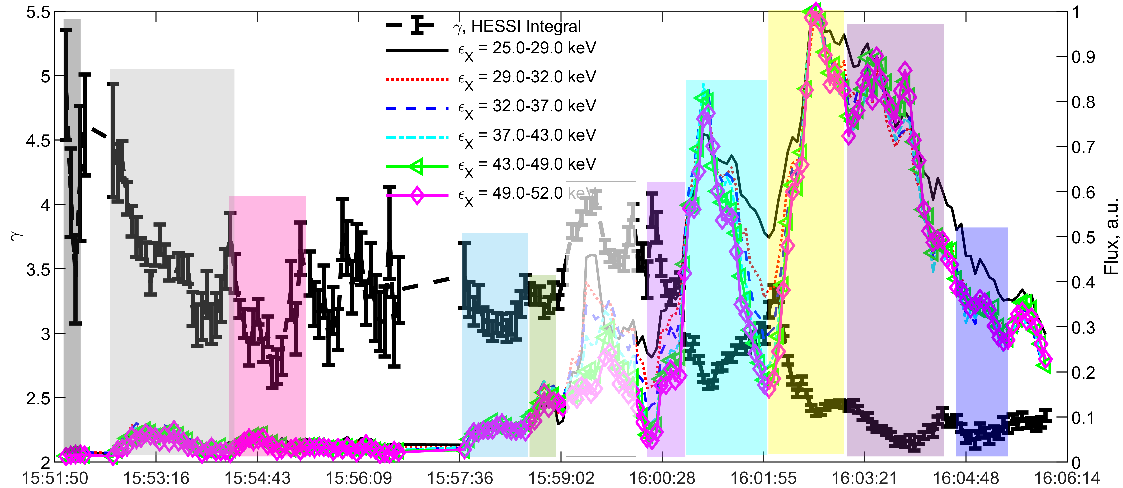}{0.90\textwidth}{}
    }
\caption{The normalized HXR fluxes and spectral indices for flares SOL2013-05-13T01:50 (top panel) and SOL2013-05-13T15:51 (bottom panel). Colored rectangles indicate the beginning and end times of the SHS patterns.
\label{fig:gamma}}
\end{figure*}

\begin{figure*}[ht!]
\centering
\includegraphics[trim=3mm 5mm 5mm 5mm, clip, width=0.95\textwidth]{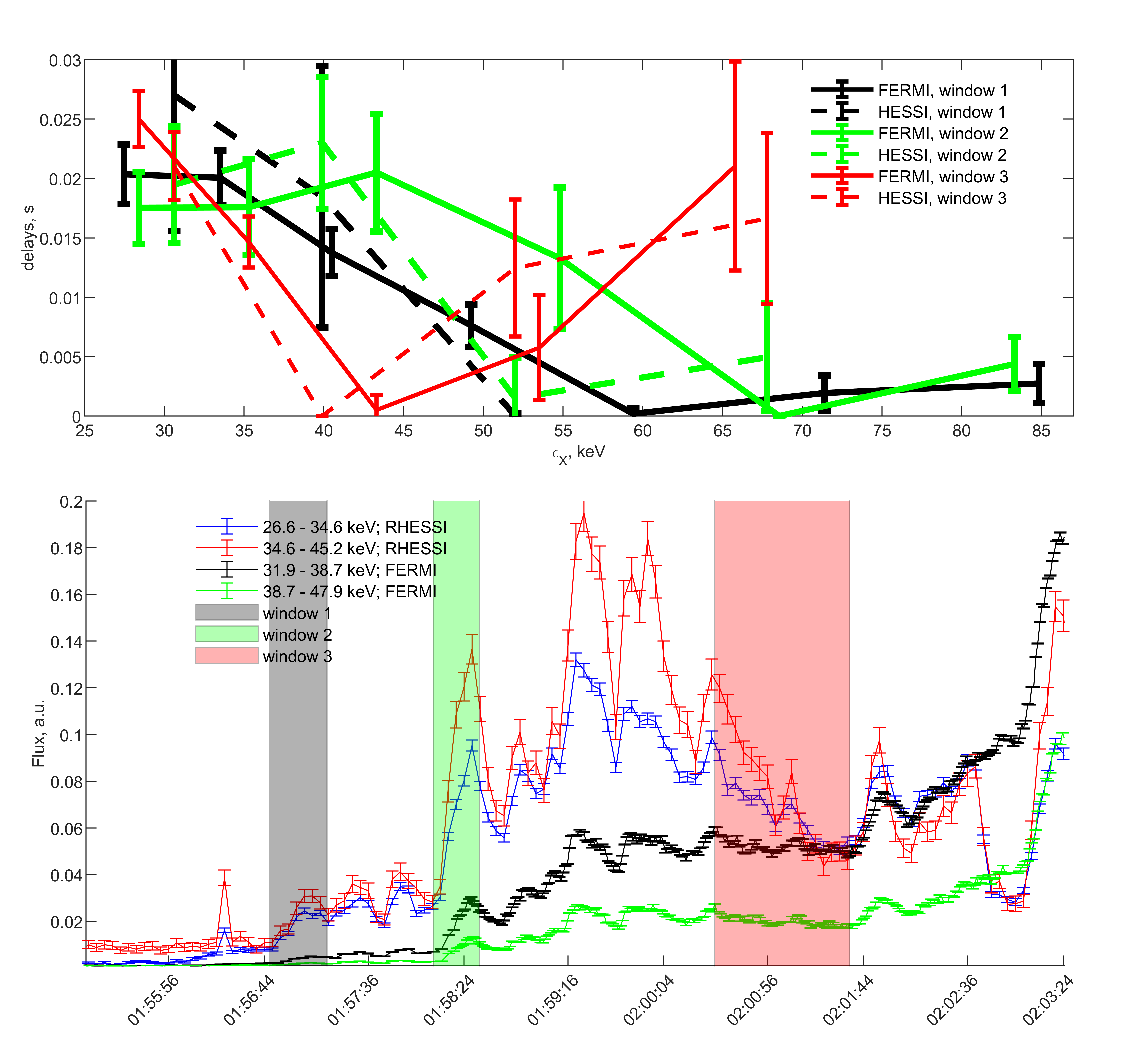}
\caption{TD spectra (top panel) and HXR fluxes (bottom panel) for the SOL2013-05-13T01:50 event obtained from GBM/Fermi and RHESSI.
\label{fig:delays_obs}}
\end{figure*}

\begin{figure*}[ht!]
\centering
\includegraphics[trim=8mm 8mm 8mm 8mm, clip, width=0.85\textwidth]{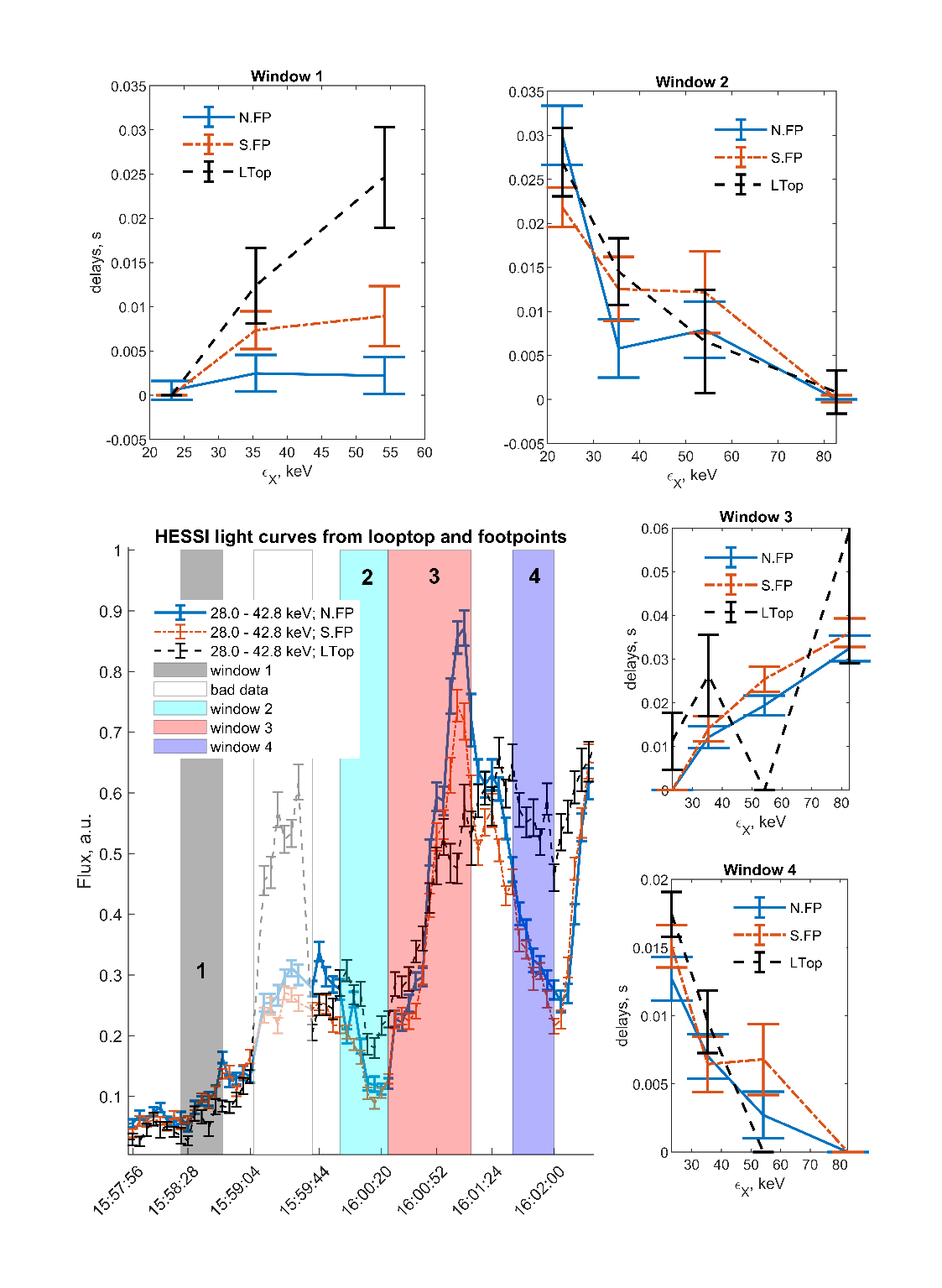}
\caption{TD spectra in time windows 1-4 and HXR fluxes (left panel) for the SOL2013-05-13T15:51 event, obtained with RHESSI in the 28.0-42.8 keV channel for HXR sources at the looptop and both footpoints of the flare arcade.
\label{fig:delays_obs_hessi}}
\end{figure*}

\section{Energy spectra of hard X-ray radiation} \label{sec:en_spec}

Figure \ref{fig:Profiles} and Table \ref{tab:table1} show the downward motion of the HXR sources during the early phases of both flares. Hereafter, the energy and time delay spectral analyses focus on this initial phase. The upper panel of Figure \ref{fig:gamma} shows the normalized HXR flux in the 46--55 keV energy range and the power-law spectrum index for the initial stage of the SOL2013-05-13T01:50 event obtained from the 4 s resolution RHESSI data. The spectral indices were derived using a two-power function approximation of the emission spectrum in the OSPEX \citep{Schwartz2002} within the 25--100 keV energy range, where the effect of the thermal plasma was insignificant. During the time interval presented in the top panel of Figure \ref{fig:gamma}, the break energy $E_{break}$ varies within the range of 38--55 keV. Hence, the spectral index corresponds to an energy range of 25--45 keV. The spectral index beyond the break energy was not provided because it was determined to be unreliable owing to insufficient photons in the coronal source at energies greater than 45 keV. At the early stage of HXR flux growth, the count rates at attenuator positions A0 and A1 are subject to the pileup effect \citep{Smith2002}. For the HXR flux shown in Figure \ref{fig:gamma}, the pileup effect was considered for the SOL2013-05-13T01:50 flare using the RHESSI software. Alternatively, the pileup effect can be incorporated into a set of fitting functions in the OSPEX software. No qualitative changes were observed for this particular event when the pileup effect was considered using the OSPEX software. The colored rectangles in Figure \ref{fig:gamma} indicate the detected SHS patterns across varying time scales ranging from 16 s to approximately 120 s. 

The lower panel of Figure \ref{fig:gamma} displays the HXR fluxes across the six energy channels, ranging from 25 keV to 52 keV, for the SOL2013-05-13T15:51 event. Analyzing the spectra from the RHESSI data for local sources with a time resolution of 4 s poses a significant challenge. This is especially the case when examining the spectrum of a coronal source in the presence of chromospheric HXR sources. Given our primary interest in the early flare stage, the analysis becomes even more complex because of the relatively small HXR flux. Therefore, the bottom panel of Figure \ref{fig:gamma} presents the spectral index derived from the full-Sun RHESSI data. Single-temperature thermal and power-law functions were employed to approximate the HXR spectra. The time interval near 15:56 was excluded from the analysis because of extremely low HXR flux. The time interval near 15:59:30 (light rectangle) was excluded from the analysis because the data were considered invalid owing to the switching of the RHESSI attenuators. The colored rectangles indicate the SHS-type dependencies between the HXR flux and the spectral index (black dashed curve). It is also worth noting that the spectral index curve exhibited a downward trend overall. As a result, the average values of the power-law spectrum index decrease, corresponding to the progressive hardening of the HXR radiation spectrum or the soft-hard-harder (SHH) pattern. 

\section{Time delay spectra of hard X-ray emission} \label{sec:delays}

Figure \ref{fig:delays_obs} shows the TD spectra for the SOL2013-05-13T01:50 event. During this event, the coronal source was predominantly observed, whereas the footpoints of the arcade remained essentially invisible (see Figure \ref{fig:images}a). Therefore, data from full-Sun observations can be used to analyze the coronal source for this event. TD spectra were derived through cross-correlation analysis of high-resolution HXR radiation across various energy channels for a coronal source. We utilized 4 s full-Sun RHESSI data (channels 26.6--34.6, 34.6--45.2, 45.5--58.9, 58.9--76.7 keV) and cspec data (detector no. 5) from the Gamma-ray Burst Monitor on the Fermi Gamma-ray Space Telescope (GBM/Fermi) \citep{Meegan2009}. The time resolution varied from 4.096 s at the onset of the flare to 1.024 s later. The GBM/Fermi data were integrated from 29 to 6 channels in the 25--90 keV energy range. In the top panel of Figure \ref{fig:delays_obs}, the energy values correspond to the midpoints of the analyzed energy ranges. To estimate the error in measuring the time delays, we simulated light curves (50 iterations) within the measurement uncertainties of the RHESSI data during the calculation process. For the GBM/Fermi data, the standard deviation within the investigated time window was considered the measurement error of the initial data. The resulting TD spectrum error for each energy channel was determined as the standard deviation for an ensemble of 50 delay values obtained for the channel under investigation. The TD spectral curve was computed as the mean of the resulting spectral ensemble. The cross-correlation function of the radiation curves in various energy ranges was approximated by using a quadratic function at three points close to its peak \citep{Aschwanden1995b}.

As an example, the bottom panel of Figure \ref{fig:delays_obs} displays the radiation curves in two energy channels: 26.6--34.6 keV and 34.6--45.2 keV, retrieved from RHESSI, and 31.9--38.7 keV and 38.7--47.9 keV, obtained from GBM/Fermi. The colored rectangles designate the time domains used to calculate the TD spectra, as illustrated in the top panel of Figure \ref{fig:delays_obs}. The solid lines in the upper panel represent the TD spectra derived from the GBM/Fermi data, and the dashed lines correspond to those obtained from the RHESSI data. The colors of the curves depicted in the top panel correspond to those of the rectangles in the bottom panel. In this study, we focused on analyzing the HXR emission from the onset of the flare to the emission peak, which occurred at 02:10:22 UT. When nonthermal electrons propagate within flare loops owing to magnetic mirroring, Coulomb, or turbulence scattering, a portion of these electrons may be captured in a magnetic trap for durations ranging from seconds to tens of seconds. The structure of the time profile of the radiation curve may comprise individual pulses superimposed on one another, representing instances of electron acceleration. To minimize the impact of these effects on the TD spectra, we selected the shortest segments of the time profile for analysis. Furthermore, we attempted to analyze the phases of the radiation rise and decay separately. The bottom panel of Figure \ref{fig:delays_obs} displays the intervals in which we obtained TD spectra with acceptable error values. These represent two phases of a local increase in the radiation flux and one phase of decay (windows 1, 2, and 3 in Figure \ref{fig:delays_obs}, bottom). A qualitative agreement can be observed when comparing the GBM/Fermi and RHESSI data (solid versus dashed curves), notwithstanding the temporal resolution and channel count differences. During the rising phases of the HXR flux (windows 1 and 2, represented by the black and green curves in Figure \ref{fig:delays_obs}, respectively), the TD spectra decreased in the approximate range of 26--70 keV. Above the inflection point in the (50--70) keV region, the TD spectra become flat. During the decay phase (window 3, red curves), the TD spectra exhibit a pronounced V-shaped profile, decreasing in the 26--44 keV range and increasing beyond $\sim$44 keV. 

Figure \ref{fig:delays_obs_hessi} shows the TD spectra obtained for the HXR of the SOL2013-05-13T15:51 event. We calculated the TD spectra for regions of interest (ROIs) in the HXR images. The images were obtained via the Clean imaging algorithm applied to the RHESSI data within the 12.0--18.3, 18.3--28.0, 28.0--42.8, 42.8--65.4, and 65.4--100 keV channels. The HXR sources from the ROIs at the looptop and two footpoints of the flare arcade were analyzed. The colored rectangles in Figure \ref{fig:delays_obs_hessi} mark the phases of the rise and decay of the flare HXR, for which the TD spectra were calculated. The shaded white time interval (15:59:04--15:59:44) was excluded from the analysis because of potentially inaccurate flux values. Windows 1 and 3 correspond to the rising phases of the HXR flux. During these time intervals, the TD spectra at the loop footpoints exhibit an increasing trend in the spectral profiles. At the looptop, the TD spectrum exhibited a growth shape (window 1, black curve) and a complex shape (window 3, black curve). During the decay phases (windows 2 and 4), the TD spectral profiles changed to a decreasing shape at the footpoints and looptop. A discussion of the theoretical justification for the TD spectral data is provided in Section \ref{sec:conclusion}.

\section{The collapsing trap model. The statement of the problem.} \label{sec:modeling}

We assume that the initial acceleration within the current sheet defines the injected electrons parameters. In the multitude of relaxing magnetic loops, we modeled one characterized by parameters representing ensemble-averaged loops that exhibit similarity. Consequently, we modeled magnetic loop relaxation accompanied by the injection of primary accelerated electrons at the looptop region.

To model the properties of the HXR mentioned above, we numerically solved the time-dependent relativistic kinetic equation for nonthermal electrons, which has the following form: \citep{hamilton1990numerical,zharkova2010diagnostics,Filatov2013,Minoshima2010}
\begin{equation}\label{eq:FP}
\begin{split}
\frac{\partial f}{\partial t} 
& = -\beta c \mu\frac{\partial f}{\partial s}+ \\
& \beta c \frac{\partial\ln(B)}{\partial s}
\frac{\partial}{\partial\mu}\left[\frac{\left(1-\mu^{2}\right)}{2}f\right]+
C_1\frac{c}{\lambda_0}\frac{\partial}{\partial E}\left(\frac{f}{\beta}\right)+ \\
& C_2\frac{c}{\lambda_0\beta^{3}\gamma^{2}}\frac{\partial}{\partial\mu}\left[\left(1-\mu^{2}\right)\frac{\partial f}{\partial\mu}\right]+ \\ 
& \frac{e\mathcal{E}\beta\mu}{m_{e}c}\frac{\partial f}{\partial E} 
+\frac{e\mathcal{E}}{m_{e}\beta c} \frac{\partial}{\partial\mu}\left[\left(1-\mu^{2}\right)f\right]+ \\ &
\frac{\partial (\dot{E}f)}{\partial E}
+\frac{\partial (\dot{\mu}f)}{\partial \mu} +S\left(E,\mu,s,t\right)
\end{split}
\end{equation}

where $f(E,\mu,s,t)$ is the non-thermal electron distribution function, $s$ is the distance along the magnetic field line ($s=0$ corresponds to the loop top), $t$ is the time, $\mu=\cos(\alpha)$ is the cosine of the pitch angle, $\lambda_{0}(s)=\frac{10^{24}}{n(s)\ln\Lambda}$, $n(s)$ -- plasma density, $\Lambda=\frac{3k_{B}T_{e}}{2e^{2}} \left( \frac{k_{B}T_{e}}{8\pi e^{2}n} \right)^{0.5} $ \citep{Ginzburg1981}, $k_{B}$ -- Boltzmann constant, $T_{e}$ -- electron temperature, $e$ -- electron charge, $\beta=vc^{-1}$, $v$ -- non-thermal electrons speed, $c$ -- speed of light, $m_{e}$ -- electron mass, $\gamma=E+1$ -- Lorentz factor of the electron, $E$ is the kinetic energy of an electron, expressed in units of the electron rest mass energy, $C_{1}=x+\frac{1-x}{2}\frac{\ln\beta^{2}g^{2}E/ \alpha_{F}^{4}}{\ln\Lambda}$, $C_{2}=\frac{1}{2}+\frac{1+g}{4}C_{1}$ \citep{leach1981impulsive,Emslie1978} -- the coefficients take into account the contribution of partially ionized plasma to the energy loss and angular scattering of fast electrons, $\alpha_{F}$ -- fine structure constant, $x$ -- fraction of ionized hydrogen atoms.
Initial plasma density at the looptop is $n^{LT}_{(t=0)}=7\times 10^{8}$ cm$^{-3}$. In the chromosphere it corresponds to hydrostatic equilibrium at $10^4$K \citep{mariska1989numerical} and is exponential between the looptop and chromosphere.  For comparison, we will also use a model with stationary magnetic field, time-independent plasma distribution and density at the looptop $n^{LT}=10^{10} \,$ cm$^{-3}$.
For a more detailed description of the terms in equation (\ref{eq:FP}) and the initial conditions, see \citet{Shabalin2022}.

If the compensation of the beam current $\textbf{\textit{j}}_{b}$ by the return current $\textbf{\textit{j}}_{rc}$ in the flaring  plasma is fulfilled, that is, $\textbf{\textit{j}}_{rc} = - \textbf{\textit{j}}_{b}$, the induced electric field must satisfy the following relation:

\begin{equation}\label{eq:SIEF}
\begin{split}
\mathcal{E}(s,t) = \frac{j_b(s,t)}{\sigma(s)} = \\
\frac{e}{\sigma(s)} \int_{E_{min}}^{E_{max}} v(E) & \,dE  \int_{-1}^{1} f(E,\mu,s,t)\mu \,d\mu
\end{split}
\end{equation}

where $\sigma(s)$ is the classical Spitzer conductivity of flare plasma determined by pair collisions of charged particles. In the case of partial plasma ionization, the conductivity $\sigma(s)$ was defined in \citet{Shabalin2022}.

The source of injected electrons was modeled as the product of four independent factors representing the distributions on the energy, pitch angle, space location, and time, that is, $S(E,\mu,s,t)=KS_{1}(E)S_{2}(\mu)S_{3}(s)S_{4}(t)$, where $K$ is a normalization constant scaled to match the chosen energy flux. The peak electron energy flux was $\sim10^{10}$ erg cm$^{-2}$s$^{-1}$, typical for M-X-class solar flares.
We consider the cases of stationary isotropic $S_{2}(\alpha)=1$ and strong anisotropic $S_{2}(\alpha)=\cos^{12}(\alpha)$ pitch-angle distributions.

The time profile of the electron injection function $S_{4}(t)$ represent a pulse followed a Gaussian function $S_{4}(t)=\exp{-\frac{(t-t_1)^2}{t_0^2}}$ centered at time $t_1=2.6$ s with a width of $t_0=1.4$ s.
The electron injection occurred spatially at the looptop, and was modeled as a Gaussian $S_{3}(s)=\exp{-\frac{(s-s_1)^2}{s_0^2}}$ centered at location $s_1=0$ cm.

The energy dependence $S_1(E)$ is represented as a broken power law with $E_{br}=200$ keV, $\delta_1=3$ or $7$ below $E_{br}$ depending on the model that corresponds to the typical flare's hard and soft spectra and $\delta_2=10$ above $E_{br}$.

The betatron and Fermi accelerations of electrons in the collTr model were calculated using the equation \citep{Filatov2013,Minoshima2010}:

\begin{equation}\label{eq:FermiBetatronE}
\dot{E}=\gamma\beta^{2}\frac{1-\mu^{2}}{2}\frac{B_{t}}{B}-\gamma\beta^{2}\mu^{2}\frac{\dot{l}}{l}
\end{equation}

where $B_t$ is the time derivative of the magnetic field. From equation (\ref{eq:FermiBetatronE}), it follows that electrons with pitch angles close to $\frac{\pi}{2}$ are mainly accelerated by the betatron mechanism, which boosts the transverse velocity component, while the longitudinal component of the electron velocity increases due to Fermi acceleration when the magnetic mirrors converge ($\dot{l}<0$). The betatron and Fermi pitch-angle terms are set as:

\begin{equation}\label{eq:FermiBetatronMu}
\dot{\mu}=-\mu\frac{1-\mu^{2}}{2}\frac{B_{t}}{B}-2\mu(1-\mu^{2})\frac{\dot{l}}{l}
\end{equation}

The magnetic field in the collTr model with a vertical current sheet cannot be strictly determined. We consider a model of the magnetic field evolution, with an initial rapid change in the magnetic field strength and length of the magnetic field lines mainly in the cusp, in the following form:

\begin{equation}\label{eq:Bst}
B(s,t)=B_{min}+\left[ \frac{s-b_{1}}{l} \right]^{2}(B_{max}-B_{min}),
\end{equation}

\begin{equation}\label{eq:Bmin}
B_{min}=B_{0}+\cos \left( \frac{2\pi t}{T_{cusp}} \right) e^{-D_{b}t} (B_{cusp}-B_{0}),
\end{equation}

where the magnetic field at the footpoints is fixed $B_{max}=mB_{0}$, the period of magnetic field oscillations $T_{cusp}=16\,$s, the initial magnetic field at the top of the loop $B_{cusp}=20$\,G, $b_1=0\,$cm defines the magnetic field minimum location, the damping factor $D_{b}=0.14\,$ s$^{-1}$, and the average magnetic field at the looptop of the relaxed magnetic loop $B_{0}=200\,$G. Thus, in the considered model, the collapse time $t_{col}$ is half of $T_{cusp}$, that is, 8 s. The coefficient $D_{b}$ is selected such that the possible following oscillations of the magnetic field dampen sufficiently quickly. The values of the magnetic field $B_{cusp}$ and $B_{0}$ were chosen as typical for the cusp region and the tops of the arcade of closed loops \citep{Gary2018,Kuridze2019}.
The  mirror ratio $m$ for closed flare loops is determined by the magnetic field gradient and varies in the interval 2--20 \citep{Shabalin2019}. We considered the results of the calculations for a symmetric magnetic loop with $m = 8$. The mirror ratio is constant for the model without collapsing traps and is the average for the model with collapsing traps. Figure 2b in \citet{Shabalin2022} shows the magnetic field variations at the looptop, according to Equation \ref{eq:Bst}.

The length of the collapsing loops during their relaxation decreases from $1.5l_0$ with a looptop in the cusp to $l_0$ for low closed loops:

\begin{equation}\label{eq:dldt}
l(t)=l_{0}-\cos \left( \frac{2\pi t}{T_{cusp}} \right) e^{-D_{b}t} (l_{1}-l_{0})
\end{equation}

where $l(t)$ is the semidistance between trap ends, $l_{0}=3\times10^{9}$ cm, $l_{1}=\frac{1}{2}l_{0}$. 

The X-ray flux was calculated according to the formulas of relativistic bremsstrahlung \citep{Bai1978,Gluckstern1953,KOCH1959} for accelerated electrons (see equation (9) in \citet{Shabalin2022}).

\section{Results of modeling the kinetics of accelerated electrons} \label{sec:modeling_res}

\subsection{Energy spectra} 
\label{subsec:modeling_E_spec}

\begin{figure*}[ht!]
\gridline{\fig{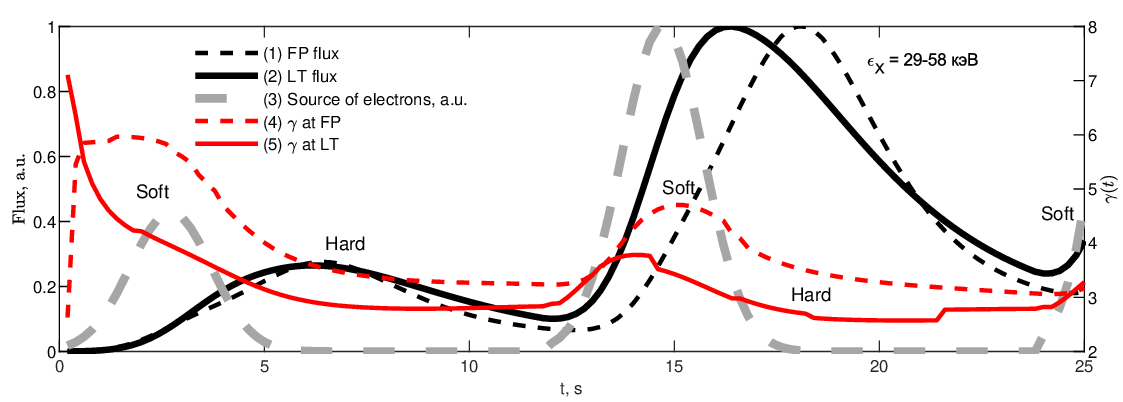}{0.9\textwidth}{}
    }
\gridline{\fig{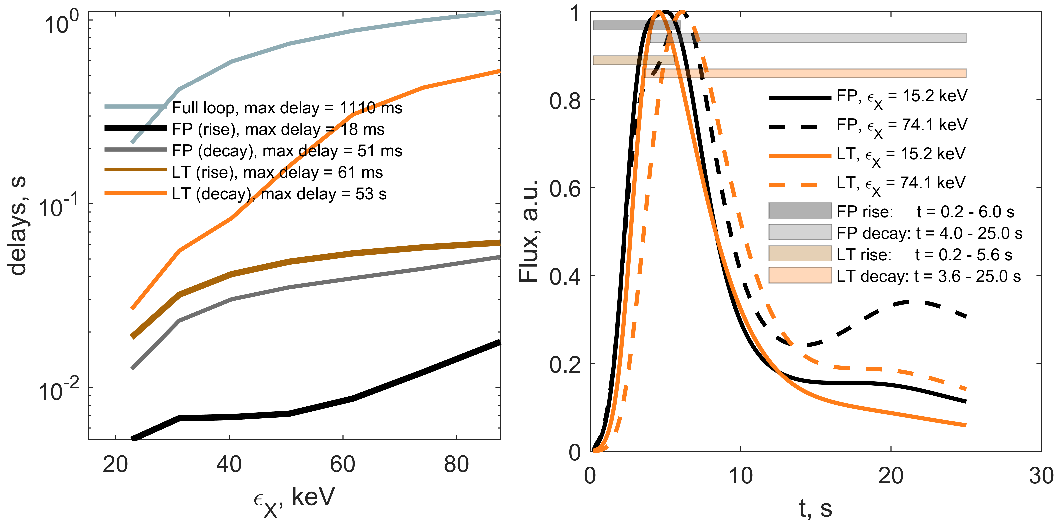}{0.7\textwidth}{}
    }
\gridline{\fig{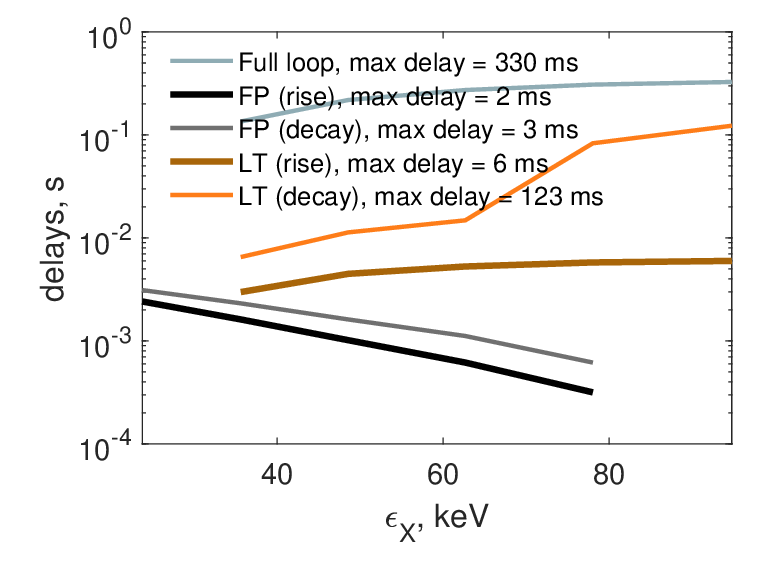}{0.43\textwidth}{}
          \fig{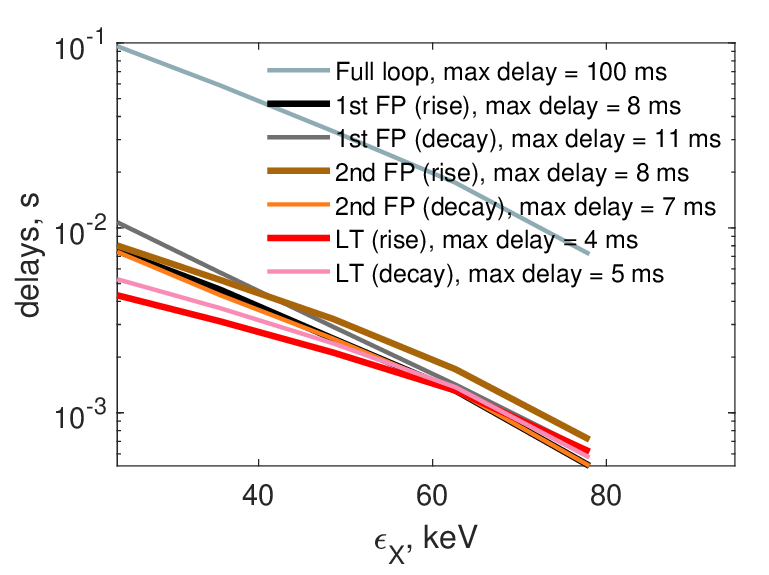}{0.44\textwidth}{}
    }
\caption{Top panel: Model HXR fluxes and spectrum index (right axis) in the 29--58 keV range for the model of sequentially collapsing traps. Middle panel: TD spectra for the CollTr model with $S(\alpha) = 1$ and $\delta = 3$. Bottom panel: On the left, TD spectra for the DMTI model with parameters $S(\alpha) = 1$, $\delta = 3$, $b_1 = 0$ cm; on the right, TD spectra for the DMTI model with parameters $S(\alpha) = 1$, $\delta = 3$, $b_1 = -2 \times 10^9$ cm, representing an asymmetric magnetic field in the flare loop. The plasma density at the looptop in the DMTI models was $10^{10}$ cm$^{-3}$.
\label{fig:modelFig}}
\end{figure*}

Magnetic field reconnection and particle acceleration can occur sporadically in time and with varying intensities as well as with different localizations within the flare region. To identify the main trends in the changes in the energy spectrum of the HXR radiation, we considered a simplified model employing successive collapse of several magnetic loops. The total injection function of the accelerated electrons for such a model appears as two Gaussian-shaped elementary bursts with different peak fluxes, which follow each other for approximately 10 s (gray dashed curve in Figure \ref{fig:modelFig}, top panel). The interval between the peaks and the full width at half maximum (FWHM) of the electron source was selected to minimize the influence of one collapsing loop (previously referred to as the CollTr model) on the subsequent loop. Additionally, the top panel displays the integral (over several collapsing loops) HXR flux in the 29--58 keV range from the looptop (solid black curve) and footpoint (dashed black curve). The red curves (solid and dashed lines) represent variations in the power-law spectrum index within the 29--58 keV energy range. The spectrum index values are shown on the right vertical axis. The labels "soft" and "hard" on the top panel indicate the formation of a soft-hard-soft pattern. During the initial increase in HXR flux, the spectrum of HXRs was soft. At the HXR peak and the onset of decay, the spectrum of the HXR radiation became harder. Furthermore, the hardness of the subsequent peaks in the spectrum index curve increased. Thus, on a time scale corresponding to the collapse of several magnetic loops, the dynamics of the spectral indices form the SHH pattern. Hence, consecutive collapses of two magnetic loops form the SHS pattern, whereas a combination of three or more collapsing loops gives rise to the SHH pattern.

\subsection{Time Delay Spectra} \label{subsec:modeling_delays}

The TD spectra for the CollTr model are shown on the left side of the middle panel in Figure \ref{fig:modelFig}. The model parameters are $S(\alpha) = 1$, $\delta = 3$, and $b_1 = 0$ cm. Following the same approach as the analysis of flares in Section \ref{sec:delays}, we calculated the model TD spectra by examining the phases of increasing HXR flux up to and including the peak, as well as the decay phases encompassing the peak. The middle panel on the right displays examples of calculated HXR fluxes at two energies, 15.2 keV and 74.1 keV, for the looptop and footpoint. It is important to note the increasing TD spectra in the CollTr model at both the footpoint and the looptop. The delays ranged from 18 to 1110 ms, depending on the analyzed spatial region of the flare loop. Alternative models were also calculated, including one with a softer electron source $S(\alpha) = 1$ and $\delta = 7$ and another with an anisotropic source $S(\alpha) = \cos^{12}(\alpha)$ and $\delta$ values of 3 and 7. Graphs for these models are not presented here. However, it is important to note that the same increasing trend in the TD spectra was observed for all calculated cases. 

To reveal the features of the CollTr model, we also calculated the TD spectra for the other models, in which the distributions of plasma density and magnetic field along the loop were time-independent (hereafter referred to as the DMTI models). In the bottom left panel of Figure \ref{fig:modelFig}, the TD spectra are presented for the DMTI model with the parameters $S(\alpha) = 1$, $\delta = 3$, and $b_1 = 0$ cm. It is important to note the decreasing TD spectra at the footpoints of the magnetic loop (black and light-gray curves). Conversely, at the looptop, the TD spectra exhibit an increasing trend. We also examined another DMTI model with the parameter $b_1 = -2\times10^9$ cm. In this model, the distribution of the magnetic field along the loop was asymmetric, featuring a $B_{max}/B_{min}$ ratio of 8 at one footpoint and 1.28 at the second one. Consequently, the geometric apex of the loop, where the majority of accelerated electrons are injected, is no longer a magnetic trap for charged nonthermal particles. The TD spectra in this model, with the parameters $S(\alpha) = 1$, $\delta = 3$, and $b_1 = -2\times10^9$ cm, are displayed in the bottom right panel of Figure \ref{fig:modelFig}. In this case, the looptop TD spectra showed a decreasing pattern. The delays span from 4 to 100 ms.

\section{Summary and discussion\label{sec:conclusion}}
\subsection{Hard X-ray source motion} \label{subsec:End_HXR_motion}

We found that the height of the coronal HXR source decreased during the early phase of flux rise in the considered flares. According to Table \ref{tab:table1}, in the selected flares SOL2013-05-13T01:50 and SOL2013-05-13T15:51, the velocities absolute values of the downward and upward motions of the HXR sources varied widely and were  (0.6 -- 23.5) and (0.6 -- 22.1) km s$^{-1}$, respectively. Note that for both flares, the HXR sources with higher energy had higher velocities for downward and upward motions (Table \ref{tab:table1}, SOL2013-05-13T01:50, intervals 1, 2; SOL2013-05-13T15:51, intervals 1, 3). A decreasing tendency in the height of the X-ray source in solar flares at the initial stage of the HXR flux rise was observed by other authors. Thus, \citet{Sui2004} recorded a decrease in the source height for three flares, and the downward motion velocities were 8--20 km s$^{-1}$, after the height decrease phase, the HXR sources moved upward at velocities of 5--40 km s$^{-1}$. A similar trend was observed in the events studied by \citet{Veronig2006}, where the average velocities of the source within the 10--30 keV energy range increased from 14 to 45 km s$^{-1}$ with increasing energy. One possible explanation for the downward motion of the HXR source may be the projection effect. That is, the heights of the arcade magnetic loops decrease along the line of sight, and sequential ignition of loops is realized. However, this interpretation contradicts the stereoscopic observations of flares in the EUV range by the Extreme Ultraviolet Imager \citep{Wuelser2004} onboard the Solar TErrestrial RElations Observatory (STEREO) spacecraft \citep{Howard2008}. These observations indicate that heating of the flare loops occurs unevenly and inconsistently throughout the arcade. Additionally, it remains unclear why the velocity of the height decrease is dependent on the energy of the HXR source, and why the downward motion of the source is observed during the initial phase of the flare only.

From another perspective, the downward movement of the HXR source at the initial flare stage was considered to be a result of the sequential relaxation of the magnetic field \citep{Sui2004,Veronig2006}. It should be noted, that during the relaxation of magnetic field, a decrease in the height of the coronal HXR source may not necessarily be observed. Certain conditions are necessary to observe a bright coronal HXR source: a high plasma density $n^{LT}>10^{10}$ cm$^{-3}$ and the ability of the trap to accumulate high-energy electrons, which implies an isotropic or quasi-transverse pitch-angle distribution of electrons after their primary acceleration in the current sheet \citep{Charikov2021,Shabalin2022}. Moreover, the spatial distribution of the magnetic field in such traps should be nearly symmetrical with the same magnetic field values at the footpoints. If we assume a symmetric magnetic field, a sufficient density of plasma and accelerated particles in the region of the magnetic field minimum can be formed shortly before the vertical motion of the relaxing magnetic loop stops. Under these conditions, a minimal decrease in the height of the HXR source may be observed. Therefore, the absence of an increase in the height of the coronal HXR source at an early stage does not contradict the CollTr model.

\subsection{Spatial separation of X-ray source centroids at different energy ranges} \label{subsec:Spatial_separation}

In our opinion, one of the most important results of observations indicating the collapse of magnetic loops in a flare is the spatial separation of X-ray sources in different energy ranges: an increase in HXR source height from lower to higher energies. Sequential involvement in the relaxation process of newly formed magnetic structures results in a configuration of magnetic loops with different heights at different energies. The nonthermal electrons in the formed earlier lower loops lose energy within seconds to tens of seconds. 
The upper bound of captured accelerated electrons lifetime with energy $\lesssim$100 keV can be estimated from the kinetic equation \ref{eq:FP} as follows $\tau_{coll}=\frac{E}{\dot{E}_{coll}} \varpropto n^{-1}E^{3/2}$ where $\dot{E}_{coll}=\frac{dE}{dt}$ determines the rate of energy loss by an electron during scattering by plasma ions. It can be seen that Coulomb loss time depends on the energy as $E^{3/2}$. Therefore, for example, $\tau_{coll}$ for a 30 keV electron in a homogeneous plasma with a density of $10^{10}$ cm$^{-3}$ will be 3.21 s, and for a 98 keV electron 17.4 s. In a flare loop plasma, as follows from equation \ref{eq:FP}, accelerated electrons, when moving along the magnetic field, pass through the loss cone into the chromosphere, scatter, reflect from magnetic mirrors, and are additionally accelerated by Fermi and betatron mechanisms. In addition, in the collTr model, at the beginning of the collapse (in a higher loop), the efficiency of the capture of accelerated electrons is high due to the large values of the magnetic ratio $m\gg 8$, and the Coulomb losses are small, because the density is low $n<10^{10}$ cm$^{-3}$. The scenario is different for relaxed low loops. The magnetic ratio in these loops is not as high, with $m \sim 5-8$, and the Coulomb losses are significantly greater because $n>10^{10}$ cm$^{-3}$.
Therefore, the lower loops radiate in a softer energy range because of hot plasma only. In magnetic loops at higher altitudes, nonthermal electrons are yet to thermalize. These electrons accumulate at the looptop due to trapping. Therefore, electron propagation in magnetic traps, combined with traps formation at different times, led to height separation of the centroids of the HXR source in various energy ranges in the considered events SOL2013-05-13T01:50 and SOL2013-05-13T15:51. Source centroid separation in various energy channels has also been observed by \citet{Veronig2006} and simulated by \citet{Kong2019}. The values of the magnetic field minima are different for relaxing magnetic loops of different heights. For higher loops, these values are lower than those for the lower loops. Therefore, the magnetic loop velocities depend on the height owing to the conservation of the magnetic flux $B\cdot v = Const$ \citep{Veronig2006}. Thus, we observed the effect of increasing the velocities with source energy for the considered flares.

\subsection{Hard X-ray flux—area correlation} \label{subsec:flux_area_correlation}

In addition to the height separation of HXR sources at different energies, we examined the temporal variation in coronal HXR source size. Owing to the conservation of magnetic flux, an increase in the magnetic field during collapse is expected to result in a reduction in the cross-sectional area of the magnetic loop. Although this value cannot be measured, it is possible to estimate the size of the HXR source associated with the loop cross-section. During the collapse, the magnetic field in the coronal part of the loop increases several times, which leads to a decrease in the cross section of the loop by the same factor. We are not expecting to observe such a significant change in the HXR source area. Nonetheless, a negative correlation may be anticipated between the cross-sectional area of the HXR sources during collapse and HXR flux. We found a negative correlation between the area of the coronal source and the HXR flux in phases with increasing flux during downward and upward motion of the sources for both considered flares (Table \ref{tab:table1}). For SOL2013-05-13T15:51, aside from the negative correlation, a positive correlation between the area and flux was observed during the phase without an intensive increase in flux within the time interval of 15:55-15:59 UT.

\subsection{Hard X-ray energy spectra variations} \label{subsec:End_HXR_spectra}

In the considered flares, we also analyzed the temporal variations in the energy spectrum. To date, numerous hypotheses have been proposed to explain the evolution of the HXR energy spectrum during solar flares. These hypotheses can be divided into four categories. First, changes in the HXR spectrum may be associated with accelerator properties or various acceleration mechanisms during several-step acceleration. Second, electron distribution functions are transformed during transport along magnetic loops via Coulomb collisions and magnetic mirroring. This transformation influences the HXR spectra. Third, the propagation of an accelerated electron beam results in the emergence of physical processes that influence the electrons responsible for initiating these processes. For example, the return current or the generation of various turbulent modes significantly modifies the distribution function of radiating electrons. Fourth, the electron propagation environment can contain MHD waves, shock fronts, and magnetic field fluctuations generated during energy release or by an external trigger. 

The types of energy spectrum evolution that correlate with HXR flux are of particular interest for research. We identified the sub-minute and minute SHS patterns in the flares under discussion. SHS patterns are challenging to explain based solely on kinetic effects during electron beam propagation. Magnetic mirroring of electrons, Coulomb losses, and return current leads to a change in the power law spectrum index in the range of 20--100 keV by fractions of units when the integral electron flux (erg cm$^{-2}$ s$^{-1}$) varies several times according to our numerical simulations. However, in the observations, the spectrum index changes by values comparable to one or more (see, for example, Figure \ref{fig:gamma}).
In this study, we interpreted SHS patterns as a superposition of several magnetic loops within the flare region. In the CollTr model, it can be assumed that each HXR pulse, ranging from seconds to tens of seconds, corresponds to a different relaxing magnetic structure. In each of these magnetic structures, accelerated electrons undergo propagation, accumulation, and emission.

Numerical simulations of electron propagation along the loop show that at the looptop, the HXR spectrum hardens over time because low-energy electrons lose energy in Coulomb collisions more efficiently. The rate of spectrum hardening depends on the plasma density in the coronal part of the loop. The plasma density at the looptop can increase in each loop because of magnetic loop relaxation and chromospheric evaporation. In addition, the response in the HXR range exhibited a delay of several seconds relative to the electron injection time in the CollTr model.

Considering the aspects mentioned above, we modeled electron propagation in the CollTr model, assuming that the collapse of different loops occurs sequentially. In a flare environment, the formation of a separate relaxing magnetic structure strongly depends on the local parameters of the plasma and magnetic field. Therefore, the duration of the injection into a separate collapsing structure and its physical parameters can vary. A simplified model of identical successive collapsing traps was suitable to demonstrate the fundamental possibility of SHS pattern formation in the HXR spectrum. The results of this simulation are shown in the top panel of Figure \ref{fig:modelFig}. As discussed in Section \ref{subsec:modeling_E_spec}, the successive relaxation of several magnetic loops results in the formation of an SHS pattern in the spectrum index dynamics. Furthermore, a global (over several HXR peaks) SHH pattern in HXR radiation is formed owing to the gradual hardening of the distribution function of the accelerated electrons, which is integrated over several magnetic loops. Hardening is caused by Coulomb losses, which transform the distribution function of the accelerated electrons in the low-energy region and due to the effective trapping of electrons in the high-energy region. 

\subsection{Hard X-ray time delays spectra} \label{subsec:End_HXR_TDspectra}

The HXR TD spectrum is an additional tool for diagnosing accelerated electrons during flares. The dynamic of the acceleration and propagation of high-energy electrons in the flare-loop plasma determines the HXR TD spectra and their evolution. In the simplest models of electron propagation, the TD spectra are determined by a power-law dependence of approximately $E^{-1/2}$ in the free-streaming model and $\sim E^{3/2}$ in the trap-plus-precipitation model \citep{Aschwanden1995b}. Due to the speed difference between electrons, high-energy electrons reach the chromosphere before lower-energy electrons. The TD spectra exhibit a descending profile in the absence of particle scattering and trapping. The processes of electron scattering and mirroring in a magnetic field lead to the trapping of electrons. These electrons spend some time in a magnetic trap before falling into the loss cone and radiating. If there is an increase in the plasma density at the loop top and/or the generation of turbulence that scatters longitudinally distributed electrons, the trapped particles will radiate more intensely in the coronal region of the loop. This results in a complex emission pattern, leading to the formation of different types of TD spectra. Numerical calculations of the propagation of accelerated electrons and bremsstrahlung HXR radiation enable a thorough analysis and identification of the factors responsible for the formation of a particular time delay spectrum.

In Section \ref{sec:delays}, for flare SOL2013-05-13T01:50, decaying TD spectra were observed in the coronal HXR source during two bursts (windows 1 and 2), according to data from both GBM/Fermi and RHESSI instruments (Figure \ref{fig:delays_obs}). During the decay phase (window 3), the TD spectrum also exhibited a decrease in the $\sim$(26--45) keV range. The observation of decreasing TD spectra at the looptop is unusual when considering the accelerated electron propagation and emission simulation.

For the second flare, SOL2013-05-13T15:51 (Figure \ref{fig:delays_obs_hessi}), a decrease in the TD spectra at the looptop was observed in time windows 2 and 4, corresponding to the local decay phases of the HXR flux. In this event, at the footpoints during the rising phases of the HXR flux (windows 1 and 3, Figure \ref{fig:delays_obs_hessi}), increasing TD spectra, which are unusual for this location, were also observed. Based on the accelerated electron propagation DMTI model, decreasing spectra are expected at the footpoints of the flare loop, as illustrated in Figure \ref{fig:modelFig} (bottom left panel). Nonetheless, in the CollTr model, the TD spectra at the footpoints exhibited growth. This is attributed to a combination of factors: a high initial $B_{max}/B_{min}$ ratio during the onset of magnetic loop relaxation, additional electron acceleration, and an increase in plasma density at the looptop. The modeling revealed that betatron acceleration has a dominant influence on the formation of growing TD spectra at footpoints. Because betatron acceleration is significant only during the growth phase of the HXR flux, the transition from increasing TD spectra at the footpoints to decreasing patterns in windows 2 and 4 in Figure \ref{fig:delays_obs_hessi} can also be explained. That is, during the increase of the HXR flux, growing TD spectra are formed at the footpoints owing to the betatron acceleration of the electrons. Subsequently, during the decay of the HXR flux, when the betatron acceleration was no longer effective and the $B_{max}/B_{min}$ ratio was not substantial (approximately 8), the TD spectra transformed to a decreasing pattern, as anticipated by the DMTI model of electron propagation with a stationary magnetic field.

Justifying the decreasing TD spectra at the looptop is more challenging. As demonstrated on the right-hand side of the bottom panel of Figure \ref{fig:modelFig}, for the model $S(\alpha)=1$, $\delta=3$, $b_1=-2\times10^9$ cm, with an asymmetric magnetic field, the TD spectra decrease at the looptop. This occurs because, in an asymmetric magnetic loop, the geometrical top of the loop does not serve as a trap for accelerated electrons. The transition from increasing TD spectra at the looptop to decreasing ones in the SOL2013-05-13T15:51 event may be due to a reduction in the $B_{max}/B_{min}$ ratio over time during magnetic field relaxation. Assuming an initial asymmetric distribution of the magnetic field along the flare loops before flare onset, a decrease in the $B_{max}/B_{min}$ ratio from large values of approximately 80 at the beginning of the relaxation (collapse) process to $B_{max}/B_{min} \simeq 8$ would lead to a more pronounced manifestation of the magnetic field asymmetry during the decay phases of the HXR flux. The results of magnetic field approximations using the nonlinear force-free field (NLFFF) method for solar disk flares suggest that asymmetry in the magnetic field distribution along magnetic loops in active regions is a relatively common occurrence. Hence, the hypothesis regarding the asymmetric magnetic field in the examined limb flares appears entirely plausible. This was further supported by the diverse brightness of the footpoints in the SOL2013-05-13T15:51 event, indirectly indicating different values for the loss cone.

\subsection{Conclusions} \label{subsec:Conclusions}

In summary, we outlined the key conclusions. Numerical simulations of accelerated electron kinetics and the associated HXR emission in the CollTr model are consistent with the observed HXR data.  First, the separation of the HXR sources by height and energy occurs naturally during magnetic field relaxation in a model with a vertical current sheet in the corona.  Second, there was a negative correlation between the area of the HXR sources in the corona and HXR flux for both flares examined in this study. Third, there were particular characteristics of the TD spectra in the investigated flares, specifically the decreasing TD spectra in the coronal source and the increasing TD spectra in the footpoints of the SOL2013-05-13T15:51 event. We explain these characteristics using numerical simulations conducted within the CollTr model framework. Fourth, our modeling of the HXR spectrum index evolution suggests that the SHS and SHH patterns may not appear due to specific processes in the accelerator or the region of suprathermal electron propagation but rather as a consequence of the superposition of HXR fluxes from various magnetic structures, which are at different stages in terms of accelerated electron accumulation and radiation. A potential extension of this research is to calculate the gyrosynchrotron radiation in similar events using the CollTr model and compare the results with observational data obtained from radio telescopes. Various turbulence modes (such as Langmuir, ion-acoustic, and whistler modes) and magnetic fluctuations may arise in secondary acceleration and electron radiation regions. Incorporating these processes into the CollTr model is another avenue of future research.

\bibliography{Shabalin2023}{}
\bibliographystyle{aasjournal}

\end{document}